\documentclass[conference,letterpaper]{IEEEtran}
\IEEEoverridecommandlockouts
\usepackage{multirow}
\usepackage{url}
\usepackage{cite}
\usepackage{amsmath,amssymb,amsfonts}
\usepackage{algorithmic}
\usepackage{graphicx}
\usepackage{textcomp}
\usepackage{censor}
\usepackage{xcolor}
\usepackage{subcaption}
\usepackage{tikz}
\usepackage{makecell}
\usepackage{tabularx}
\usepackage{array}
\usepackage{enumitem}
\usepackage{balance}
\usepackage{booktabs}
\usepackage{amssymb}
\usepackage[table]{xcolor}
\usetikzlibrary{positioning, fit, backgrounds, arrows.meta, calc}
\usepackage{etoolbox}
\usepackage{caption}
\usepackage{tikz}
\usepackage{fancyhdr}

\title{5G-NetBotSim: A Synchronized Co-Simulation Framework for Connected Robots over ROS and srsRAN-based 5G Network}

\author{\IEEEauthorblockN{
\large Narges Golmohammadi, Bhaskara Venkata Raju Garuda, Ayan Joshi, and Sabur Baidya\\
\normalsize Department of Computer Science and Engineering, University of Louisville, KY, USA}
\normalsize {e-mail:  n0golm01@louisville.edu, b0garu01@louisville.edu, ayan.joshi@louisville.edu, sabur.baidya@louisville.edu}
}

\newcommand{\myteaserimage}{
  \noindent
  \begin{minipage}{\textwidth}
    \centering
    \vspace{2mm}
    \includegraphics[width=0.8\linewidth]{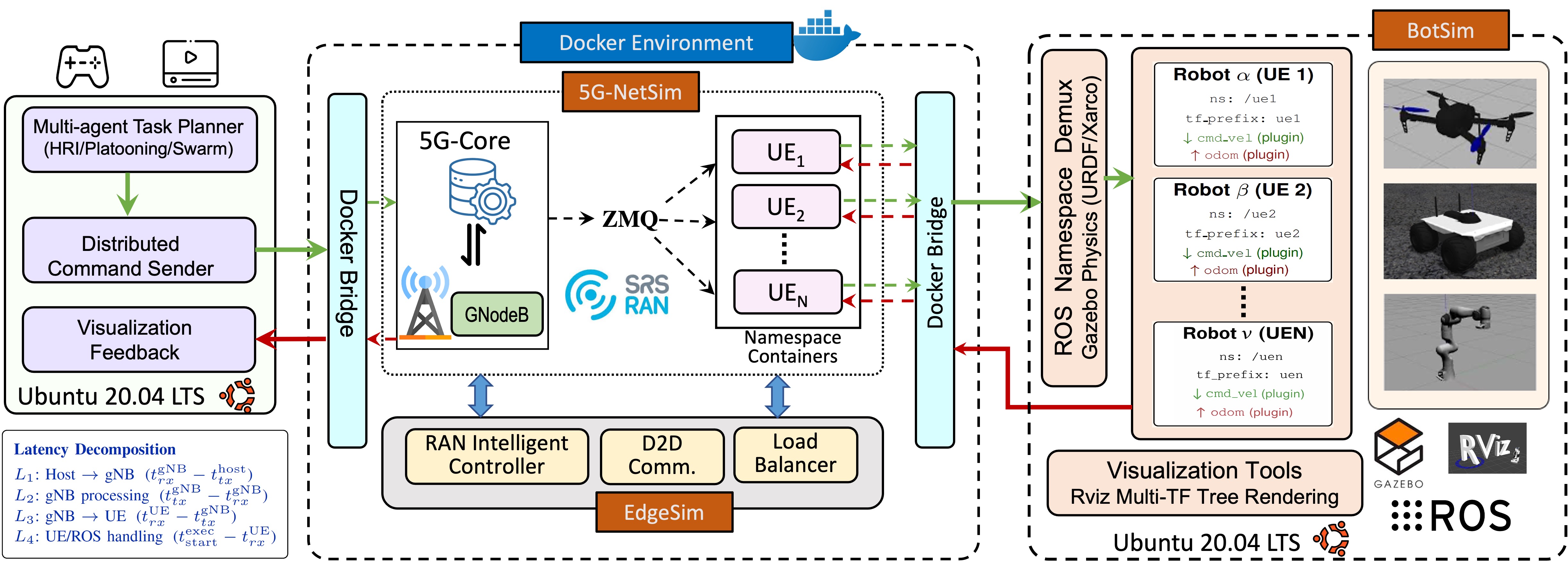}\\
    \makeatletter
    \def\@captype{figure} % Temporarily tells LaTeX this is a figure float
    \makeatother
    {\centering{\small Fig. 1.~~5G-NetBotSim architecture integrating an executable srsRAN-based 5G with isolated ROS/Gazebo robotic applications.}}
    \label{fig:architecture}
  \end{minipage}
  \bigskip
  \vspace{-8mm}
}

\makeatletter
\apptocmd{\@maketitle}{\myteaserimage}{}{}
\makeatother

\begin{document}

\maketitle
\thispagestyle{fancy}   % removes page number on page 1
\pagestyle{fancy} 

\fancyhf{}
\pagestyle{fancy}
\thispagestyle{fancy}
\renewcommand{\headrulewidth}{0pt}
\renewcommand{\footrulewidth}{0pt}

\setlength{\headheight}{14pt}
\setlength{\headsep}{10pt}

\fancyhead[C]{%
\footnotesize\makebox[\textwidth][c]{This paper has been accepted as a poster paper at ACM MSWiM 2026.}}

% FOOTER (unchanged)
\fancyfoot[C]{
    \begin{tikzpicture}[remember picture, overlay]
        \node[anchor=south,yshift=10pt] at (current page.south) {
            \parbox{\textwidth}{
                \centering
                \footnotesize
                \textcopyright~2026 IEEE. Personal use of this material is permitted. Permission from IEEE must be obtained for all other uses, in any current or future media, including reprinting/republishing this material for advertising or promotional purposes, creating new collective works, for resale or redistribution to servers or lists, or reuse of any copyrighted component of this work in other works.
            }
        };
    \end{tikzpicture}
}

\begin{abstract}
Connected robotic applications increasingly rely on wireless communication for teleoperation, coordination, and distributed autonomy. However, robotic simulators often use ideal links or abstract delay models that do not capture cellular scheduling and protocol-processing effects. We present \textit{5G-NetBotSim}, a real-time co-simulation framework integrating ROS/Gazebo workloads with an executable, full-stack srsRAN 5G network emulator. The framework provides isolated User Equipment (UE)--robot contexts and stage-level timing across the host, gNB, UE, ROS, and robot-execution stages. We evaluate human-to-robot teleoperation and robot-to-robot leader--follower coordination under controlled channel-quality conditions. Results show how CQI and MCS variations propagate to throughput, communication latency, and robot performance. In leader--follower coordination, higher channel quality reduces mean tracking error from 0.558~m to 0.460~m, a 17.6\% reduction. These results demonstrate protocol-stack-in-the-loop co-simulation for joint network and robotic-application evaluation.
\end{abstract}


\vspace{-4mm}
\section{Introduction}

Connected robotic systems increasingly rely on wireless communication for teleoperation, multi-robot coordination, distributed perception, and collaborative autonomy. In such applications, delay, jitter, throughput variation, and packet loss can directly affect robot response, trajectory accuracy, coordination, and task completion.
Robotic simulators typically model dynamics, sensing, planning, and control in detail, but often represent communication using ideal links or application-level delay models. These abstractions do not capture cellular scheduling, modulation and coding adaptation, protocol processing, per-device isolation, or RAN configuration. Conversely, cellular-network platforms mainly report network-level metrics without integrating the complete ROS-enabled command, execution, and feedback path. This makes it difficult to determine how network conditions propagate to robot-level performance.
%To address this gap, we present \textit{5G-NetBotSim}, a synchronized protocol-stack-in-the-loop co-simulation framework that integrates ROS/Gazebo robotic workloads with an executable srsRAN 5G stack. Rather than replacing communication with a fixed or stochastic delay model, robot messages traverse the cellular protocol-processing path, including PHY, MAC, RLC, PDCP, and higher-layer components. Stage-level instrumentation then relates network behavior directly to robotic execution.

To address this gap, we present \textbf{5G-NetBotSim}, a synchronized protocol-stack-in-the-loop co-simulation framework that integrates ROS/Gazebo robotic workloads with an executable, standards-aligned srsRAN 5G stack, shown in Fig.~\ref{fig:architecture}. 
The framework executes the complete 5G protocol-processing path, including physical (PHY), medium access control (MAC), radio link control (RLC), packet data convergence protocol (PDCP), and higher-layer components.
thereby capturing cellular scheduling, protocol processing, and channel-dependent communication behavior within the robotic control loop. Rather than replacing the network with a fixed or stochastic delay model, 5G-NetBotSim preserves the full protocol-processing path and instruments each stage to relate network metrics directly to task-facing behavior.

\newcommand{\yes}{\checkmark}
\newcommand{\no}{--}

\begin{table*}[!t]
\centering

\scriptsize
\setlength{\tabcolsep}{7.5pt}
\renewcommand{\arraystretch}{1.2}

\resizebox{\textwidth}{!}{%
\begin{tabular}{lccccccccc}
\toprule
\textbf{Framework} &
\makecell{\textbf{Robot--network}\\\textbf{co-simulation}} &
\makecell{\textbf{Multi-robot}\\\textbf{support}} &
\makecell{\textbf{5G/RAN}\\\textbf{aware}} &
\makecell{\textbf{Executable}\\\textbf{5G stack}} &
\makecell{\textbf{Stage-level}\\\textbf{timing}} &
\makecell{\textbf{Robot-task}\\\textbf{metrics}} &
\makecell{\textbf{Per-robot}\\\textbf{isolation}} &
\makecell{\textbf{RAN-edge}\\\textbf{integration}} &
\makecell{\textbf{D2D MAC}\\\textbf{bypass}} \\
\midrule

CORNET~\cite{acharya2020cornet}
    & \yes & \no  & \no  & \no  & \no  & \yes & \no  & \no  & \no \\

SynchroSim~\cite{dey2022synchrosim}
    & \yes & \yes & \no  & \no  & \no  & \yes & \no  & \no  & \no \\

OROS~\cite{romero2023oros}
    & \yes & \yes & \yes & \no  & \no  & \yes & \no  & \yes & \no \\

FogROS2 family~\cite{chen2023fogros2,chen2024fogros2,ichnowski2023fogros2}
    & \no  & \yes & \no  & \no  & \no  & \yes & \no  & \yes & \no \\

NetROS-5G~\cite{dalgkitsis2023netros}
    & \yes & \no  & \yes & \no  & \no  & \yes & \no  & \yes & \no \\

5G Network Digital Twin~\cite{roda2025network}
    & \yes & \no  & \yes & \no  & \no  & \yes & \no  & \no  & \no \\

EdgeRIC~\cite{ko2023edgeric}
    & \no  & \no  & \yes & \yes & \no  & \no  & \no  & \yes & \no \\

\rowcolor{gray!18}
\textbf{5G-NetBotSim}
    & \textbf{\yes}
    & \textbf{\yes}$^{\dagger}$
    & \textbf{\yes}
    & \textbf{\yes}
    & \textbf{\yes}
    & \textbf{\yes}
    & \textbf{\yes}
    & \textbf{\yes}
    & \textbf{\yes} \\

\bottomrule
\multicolumn{10}{l}{
\scriptsize $^{\dagger}$Multi-robot execution is experimentally
demonstrated with two independently isolated robot--UE instances; Future scope involves more experiments with a swarm of a larger number of robots.}
\end{tabular}%
}
\vspace{-2mm}
\caption{Capability comparison with representative robot-network,
edge-robotics, and RAN-control platforms.}
\label{tab:related_work}

\vspace{-6mm}
\end{table*}

5G-NetBotSim consists of three integrated components:
\textit{5G-NetSim}, which provides the virtual-radio 5G communication path using the executable srsRAN protocol stack;
\textit{BotSim}, which provides ROS/Gazebo-based robot simulation with separate ROS and network contexts for individual robots; and
\textit{EdgeSim}, which extends the RAN-side architecture with edge intelligence using EdgeRIC.

The main contributions are:
\begin{itemize}[leftmargin=2mm] \itemsep 1pt
    \item We develop a synchronized protocol-stack-in-the-loop framework coupling ROS/Gazebo with an executable srsRAN 5G stack.
    
    \item We provide per-robot UE, network, and ROS isolation with sequence-aligned timing across host, gNB, UE, ROS, and robot-execution stages.
    
    \item We integrate EdgeRIC through EdgeSim for RAN-side intelligent control and communication-aware robotic applications.
    
    \item We evaluate teleoperation and leader--follower coordination under controlled CQI regimes, linking network conditions to robot performance.
\end{itemize}

\iffalse

\begin{figure*}[t]
    \centering
    \includegraphics[width=0.8\linewidth]{images/5g-netbotsim.jpg}
    \vspace{-2mm}
    \caption{5G-NetBotSim architecture integrating an executable srsRAN-based 5G with isolated ROS/Gazebo robotic applications.}
    \label{fig:architecture}
    \vspace{-6mm}
\end{figure*}

\fi

\section{Related Work}
\label{sec:related_work}
Prior robot--network co-simulation frameworks such as CORNET and SynchroSim jointly evaluate robotic behavior and communication-network performance~\cite{acharya2020cornet,dey2022synchrosim}. However, these platforms primarily focus on general network modeling and simulation synchronization rather than executing a cellular protocol stack within the complete robotic command and execution path.
Several studies investigate 5G-enabled robotics, network orchestration, and cloud/edge support. OROS jointly manages collaborative robot operation and wireless resources~\cite{romero2023oros}, while task-aware 5G resource-allocation approaches improve communication and control performance in latency-sensitive telerobotic applications~\cite{golmohammadi2025hierarchical,golmohammadi2024lyapunov,golmohammadi2026drastic}. NetROS-5G integrates ROS~2 with network slicing and edge/cloud resources~\cite{dalgkitsis2023netros}, network digital-twin approaches model communication-aware 5G-connected robots~\cite{roda2025network}, and FogROS2 supports cloud/fog deployment of robotic components~\cite{ichnowski2023fogros2,chen2023fogros2,chen2024fogros2}. EdgeRIC further enables near-real-time application-aware RAN control~\cite{ko2023edgeric}.
Communication-aware robotic studies show that network delay and middleware behavior can directly affect teleoperation and multi-robot performance~\cite{farajiparvar2020brief,rakita2020effects}. Motivated by these observations, 5G-NetBotSim integrates ROS/Gazebo workloads with an executable srsRAN 5G stack and EdgeRIC-based RAN intelligence, while providing isolated UE--robot contexts and stage-level timing across the network, middleware, and robot-execution stages.

\section{System Design and Architecture}

The 5G-NetBotSim integrates a 5G environment into ROS/Gazebo applications. Each robot is mapped to an independent UE, Linux namespace, ROS namespace, and control process. Commands are transferred through ZeroMQ to the Dockerized gNB, traverse the executable 5G downlink, and are forwarded from the selected UE to the corresponding ROS process. Sequence identifiers and monotonic timestamps align network and robot-execution events.

\subsection{5G-NetSim: Executable 5G Network Emulation}

5G-NetSim uses the simulation mode of the open-source srsRAN stack. The gNB and UE processes run inside Docker and retain PHY, MAC, RLC, PDCP, and higher-layer processing through a virtual-radio interface, enabling repeatable network configuration.
Each UE operates in an independent network context mapped to a robot, supporting per-device routing, configuration, logging, and measurements such as CQI, MCS, throughput, and packet latency. The experiments use virtual-radio rather than over-the-air RF transmission, focusing on controlled protocol-stack execution and its impact on robotic applications.

\subsection{BotSim: Isolated Robotic Execution}
\label{subsec:botsim}

BotSim provides the robotic application layer using ROS Noetic and Gazebo 11. Each UE endpoint connects to a dedicated ROS namespace and robot instance, forwarding received commands to the corresponding ROS control node for planning or execution.

The one-to-one UE-to-robot mapping provides routing and measurement isolation. ZeroMQ is used only at the host--Docker boundaries to keep ROS/Gazebo and srsRAN independently configurable and does not replace the evaluated gNB--UE path. This separation also allows robotic workloads and network configurations to be modified independently.

\subsection{EdgeSim: RAN-Side Edge Integration}

EdgeSim provides RAN-side support for edge-assisted communication and control using EdgeRIC, enabling near-real-time interaction with the RAN and application-aware network control without modifying the ROS/Gazebo layer.

\subsection{End-to-End Integration and Timing Instrumentation}
\label{subsec:integration_timing}

Each command $k$ carries a sequence identifier and timestamps generated using \texttt{time.monotonic\_ns()}. Because the Dockerized 5G and host-side ROS processes run on the same computer, they share a common monotonic timing reference.
The stage-level delays are
\begin{equation}
\vspace{-2mm}
\begin{aligned}
L_1 &= t_{\mathrm{rx}}^{\mathrm{gNB}}-t_{\mathrm{tx}}^{\mathrm{app}}, &
L_2 &= t_{\mathrm{tx}}^{\mathrm{gNB}}-t_{\mathrm{rx}}^{\mathrm{gNB}},\\
L_3 &= t_{\mathrm{rx}}^{\mathrm{UE}}-t_{\mathrm{tx}}^{\mathrm{gNB}}, &
L_4 &= t_{\mathrm{start}}^{\mathrm{exec}}-t_{\mathrm{rx}}^{\mathrm{UE}},\\
L_5 &= t_{\mathrm{done}}^{\mathrm{exec}}-t_{\mathrm{start}}^{\mathrm{exec}}.
\end{aligned}
\label{eq:latency_components}
\end{equation}
Here, $L_1$--$L_5$ represent host-to-gNB transfer, gNB processing, gNB-to-UE communication, UE-to-ROS handoff, and robot execution, respectively. Thus,
\begin{equation}
L_{\mathrm{delivery}} = L_1 + L_2 + L_3.
\end{equation}
This instrumentation separates communication delay from application-side queueing, planning, and execution. The single-host setup also avoids cross-machine clock-synchronization uncertainty.
All experiments run on Ubuntu 20.04 with one gNB, two UEs, 10~MHz bandwidth, 15~kHz subcarrier spacing, 1~ms TTI, and a MAX-CQI scheduler. Low-, medium-, and high-CQI regimes are evaluated with fixed robotic workloads and control parameters.

\section{Performance Evaluation}

This section evaluates whether the configured CQI conditions propagate consistently through the executable 5G stack to produce measurable effects at the network and application levels, from MCS selection and throughput to communication latency and robot-task performance.

%==============================

\subsection{Network-Level Performance}

\begin{figure}[t]
\centering

\begin{subfigure}[t]{0.49\linewidth}
    \centering
    \includegraphics[width=\linewidth]{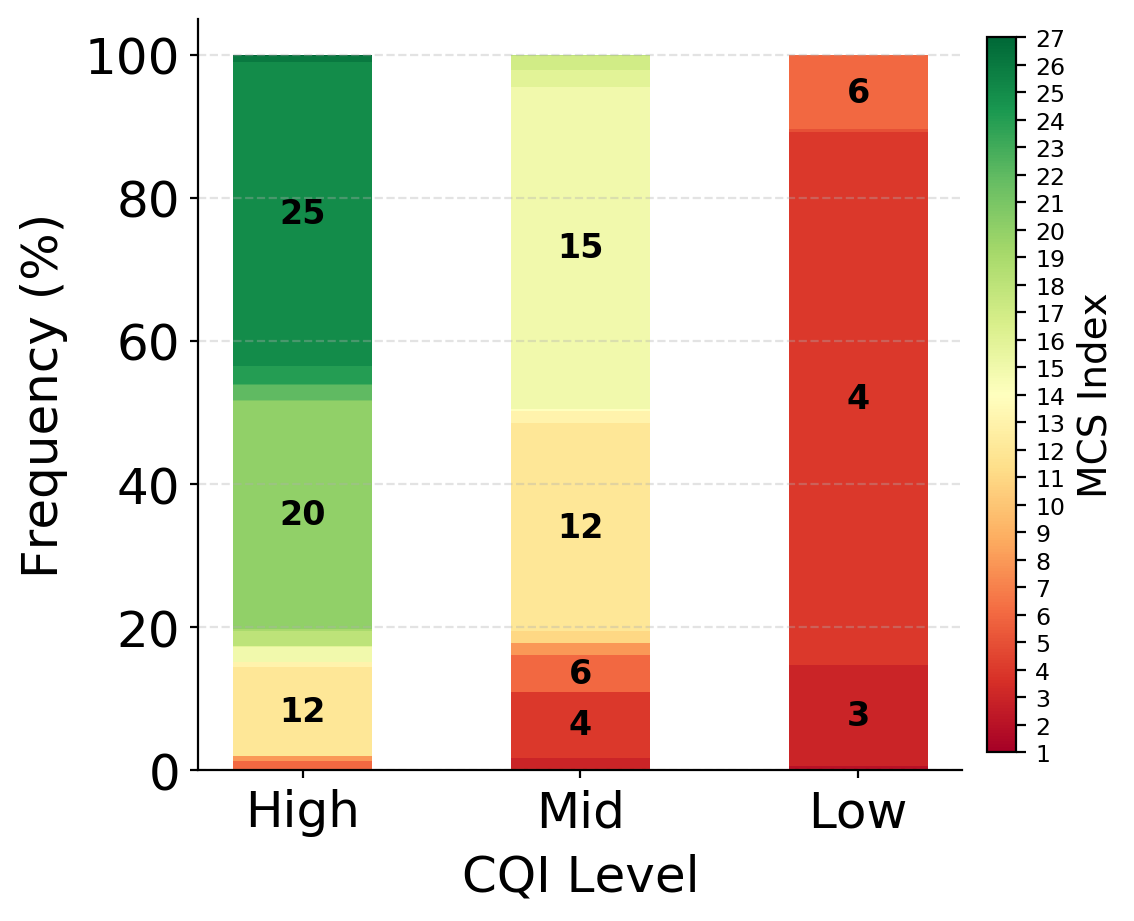}
    \caption{Downlink MCS}
    \label{fig:mcs_dl_bar}
\end{subfigure}
\hfill
\begin{subfigure}[t]{0.49\linewidth}
    \centering
    \includegraphics[width=\linewidth]{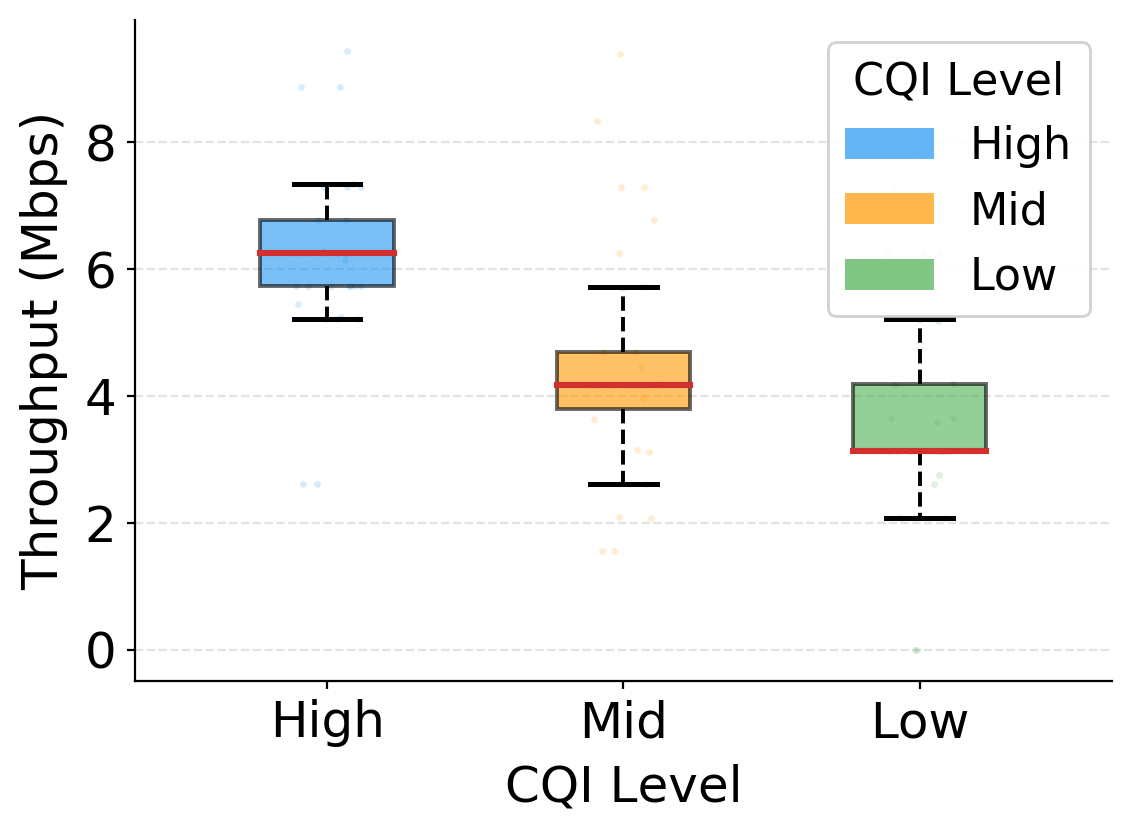}
    \caption{Throughput measurements}
    \label{fig:throughput_whisker}
\end{subfigure}

\vspace{-1mm}
\caption{Selected MCS for different CQI levels and corresponding throughput results.}
\label{fig:mcs_throughput}
\vspace{-5mm}
\end{figure}

\begin{figure*}[!t]
\centering

\begin{subfigure}[t]{0.29\textwidth}
    \centering
    \includegraphics[
        width=\linewidth,
        height=2.8cm,
        keepaspectratio
    ]{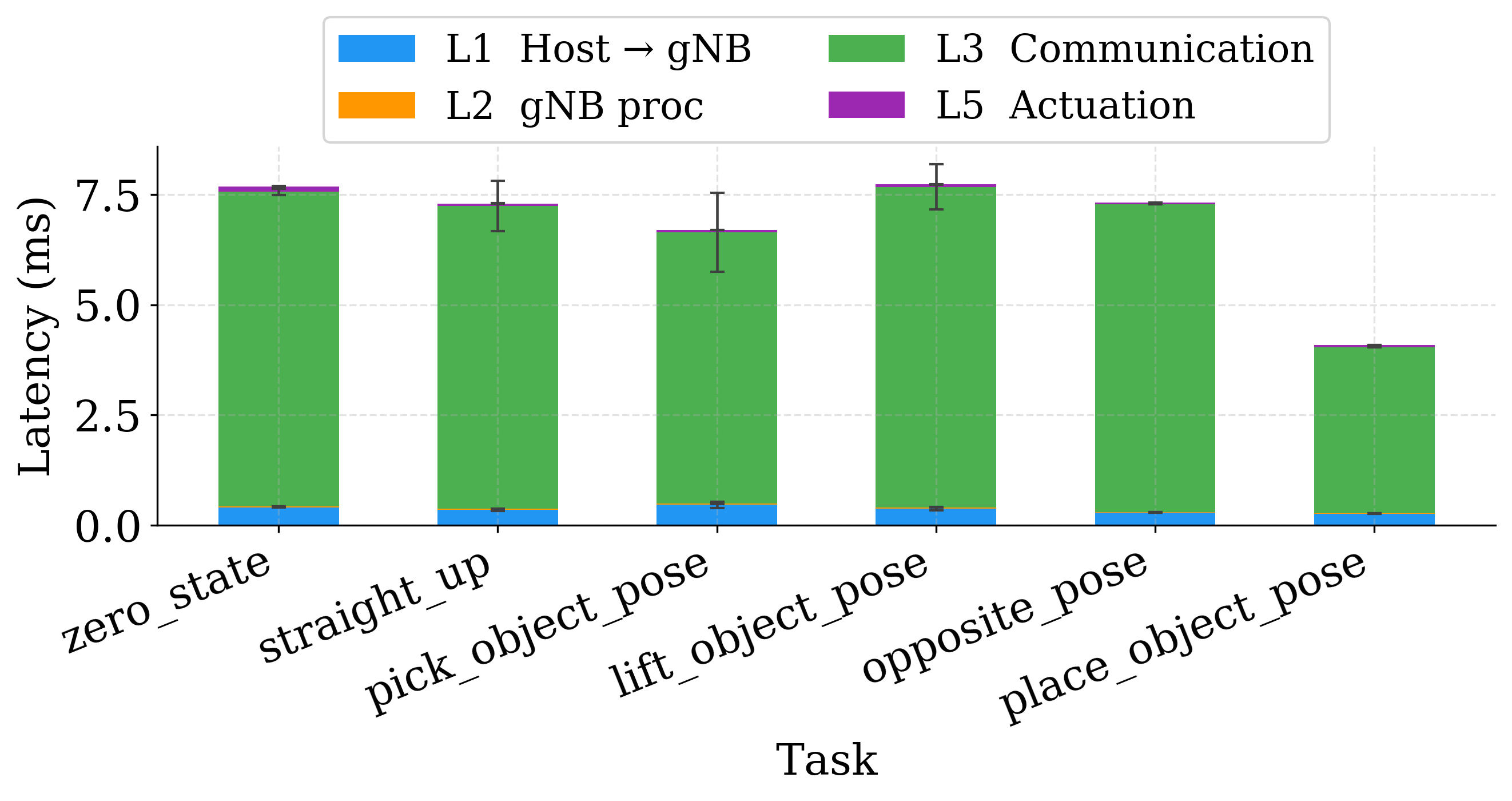}
    \caption{E2E latency}
    \label{fig:e2e_latency}
\end{subfigure}
\hfill
\begin{subfigure}[t]{0.31\textwidth}
    \centering
    \includegraphics[
        width=\linewidth,
        height=2.8cm,
        keepaspectratio
    ]{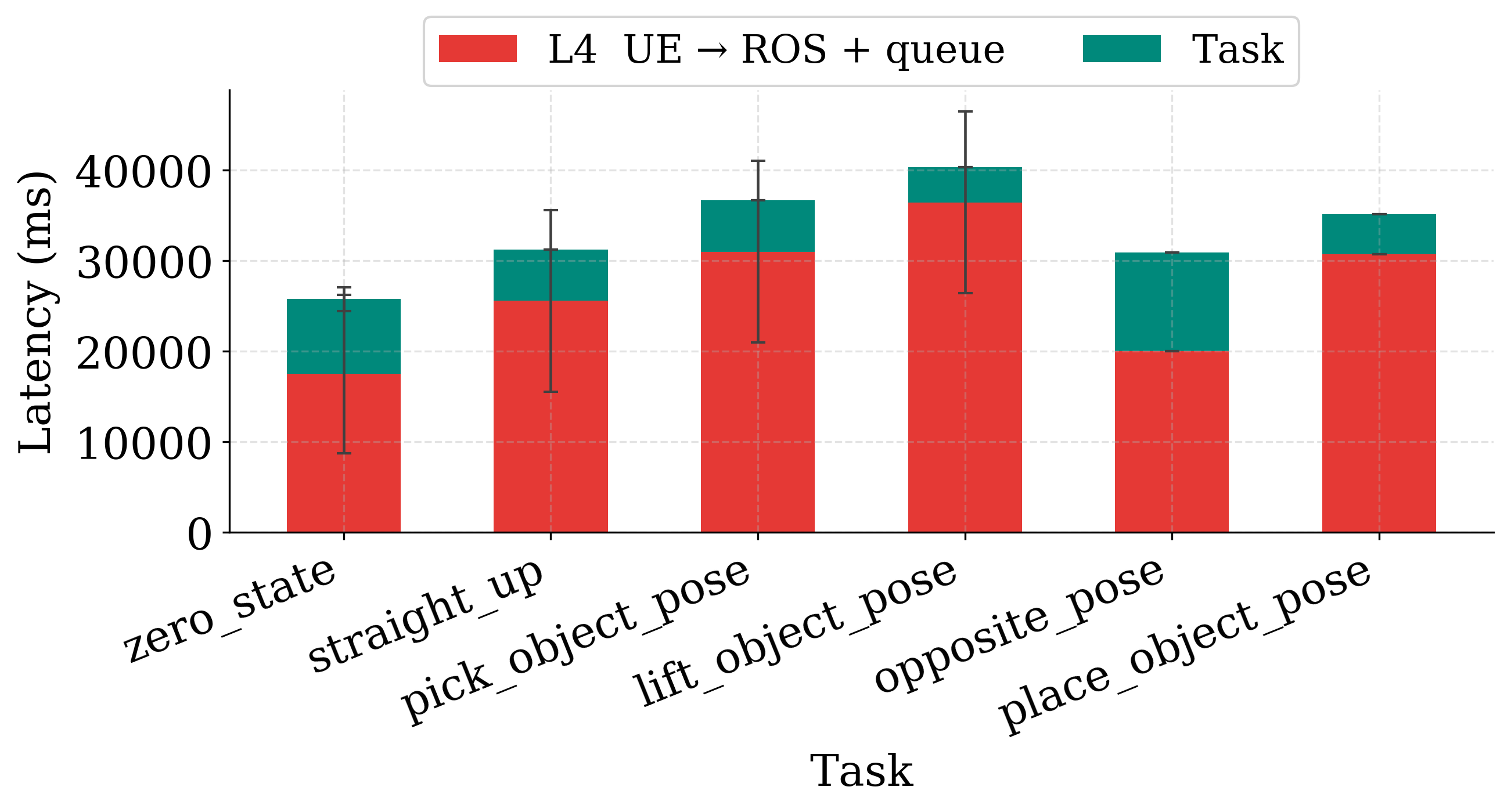}
    \caption{Task latency}
    \label{fig:task_latency}
\end{subfigure}
\hfill
\begin{subfigure}[t]{0.29\textwidth}
    \centering
    \includegraphics[
        width=\linewidth,
        height=2.8cm,
        keepaspectratio
    ]{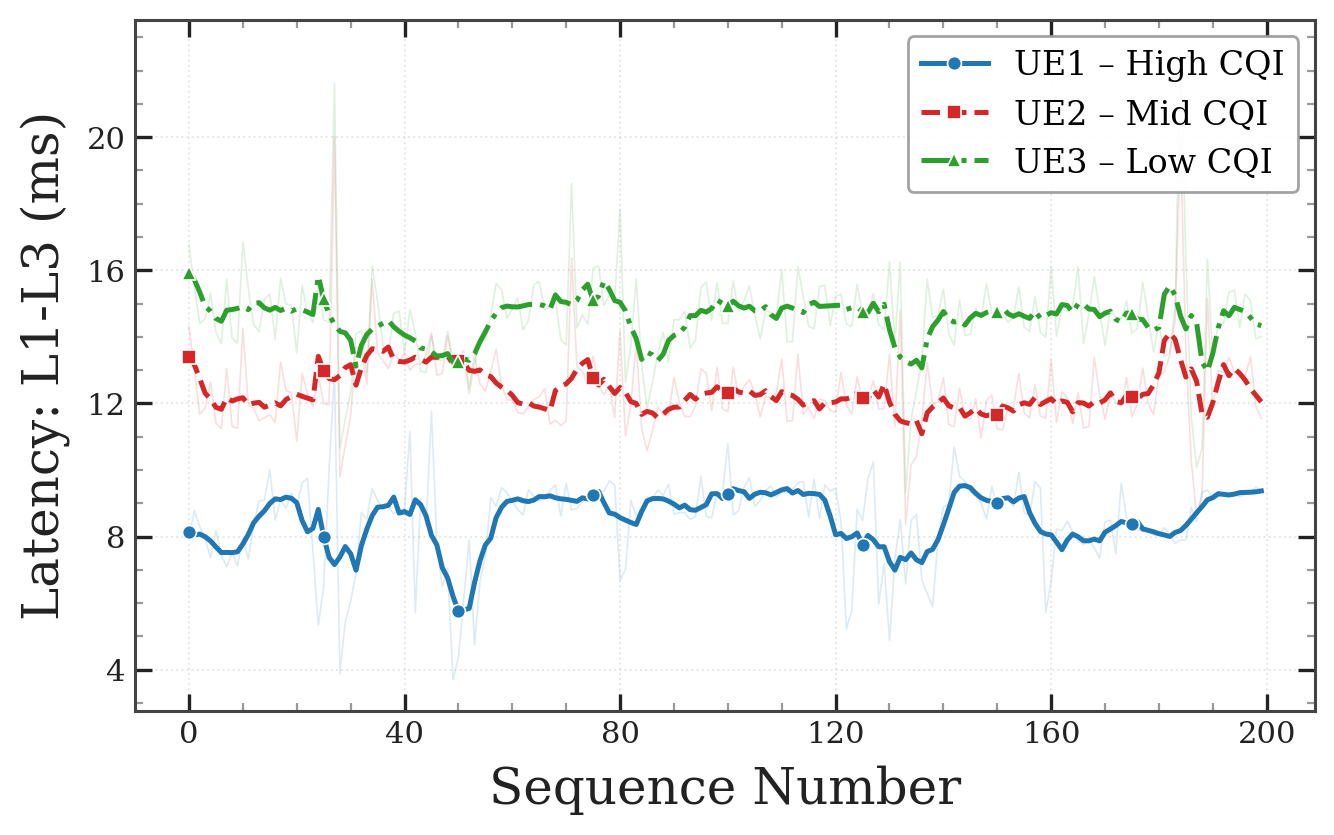}
    \caption{Total latency}
    \label{fig:latency_compare_total}
\end{subfigure}

\vspace{-2mm}
\caption{Application- and task-level latency measurements.}
\label{fig:latency_combined_top}
\vspace{-4mm}
\end{figure*}

Fig.~\ref{fig:mcs_throughput} shows the selected MCS and corresponding user-plane throughput under low-, medium-, and high-CQI conditions. Higher CQI results in higher MCS indices and greater achievable throughput, while lower CQI leads to more robust modulation and coding and reduced data rate. These results confirm that the configured CQI conditions affect the actual scheduling and transmission behavior of the executable 5G stack rather than remaining only simulation parameters.
The stage-level latency measurements, shown in Fig.~\ref{fig:latency_combined_top}, further decompose command delivery across the host-to-gNB transfer, gNB processing, gNB-to-UE transmission, and robot-side handling. This decomposition helps distinguish network delay from application-side processing and execution effects, which is important because low communication latency alone does not necessarily imply low task-facing response time.

\subsection{Application-Level Performance}
\begin{figure}[t]
\centering

\begin{subfigure}[t]{0.75\linewidth}
    \centering
    \includegraphics[
        width=\linewidth,
        keepaspectratio
    ]{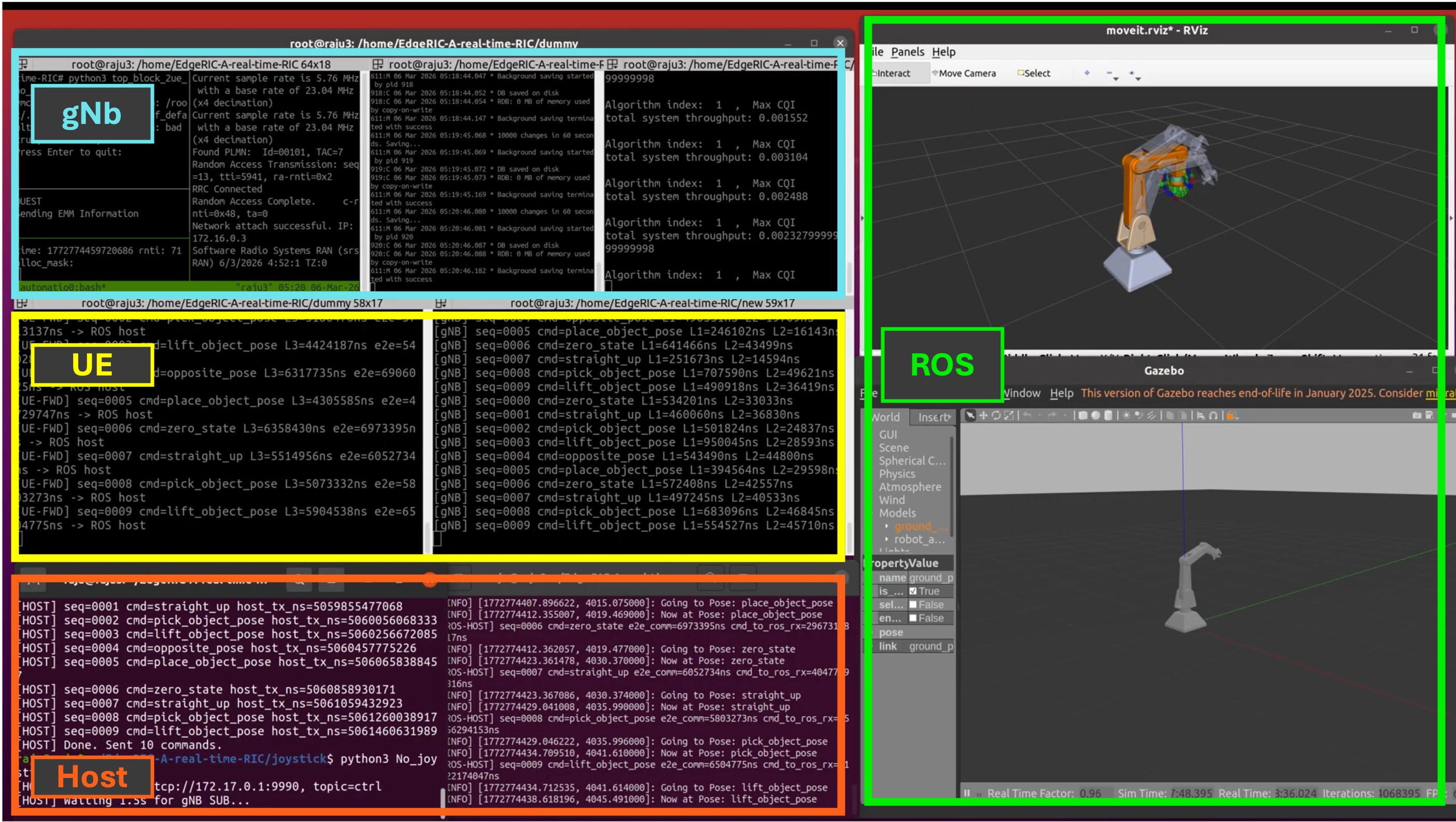}
    \caption{Human-to-Robot Teleoperation}
    \label{fig:human-robot}
\end{subfigure}

\vspace{1mm}

\begin{subfigure}[t]{\linewidth}
    \centering
    \includegraphics[
        width=0.4\linewidth,
        keepaspectratio
    ]{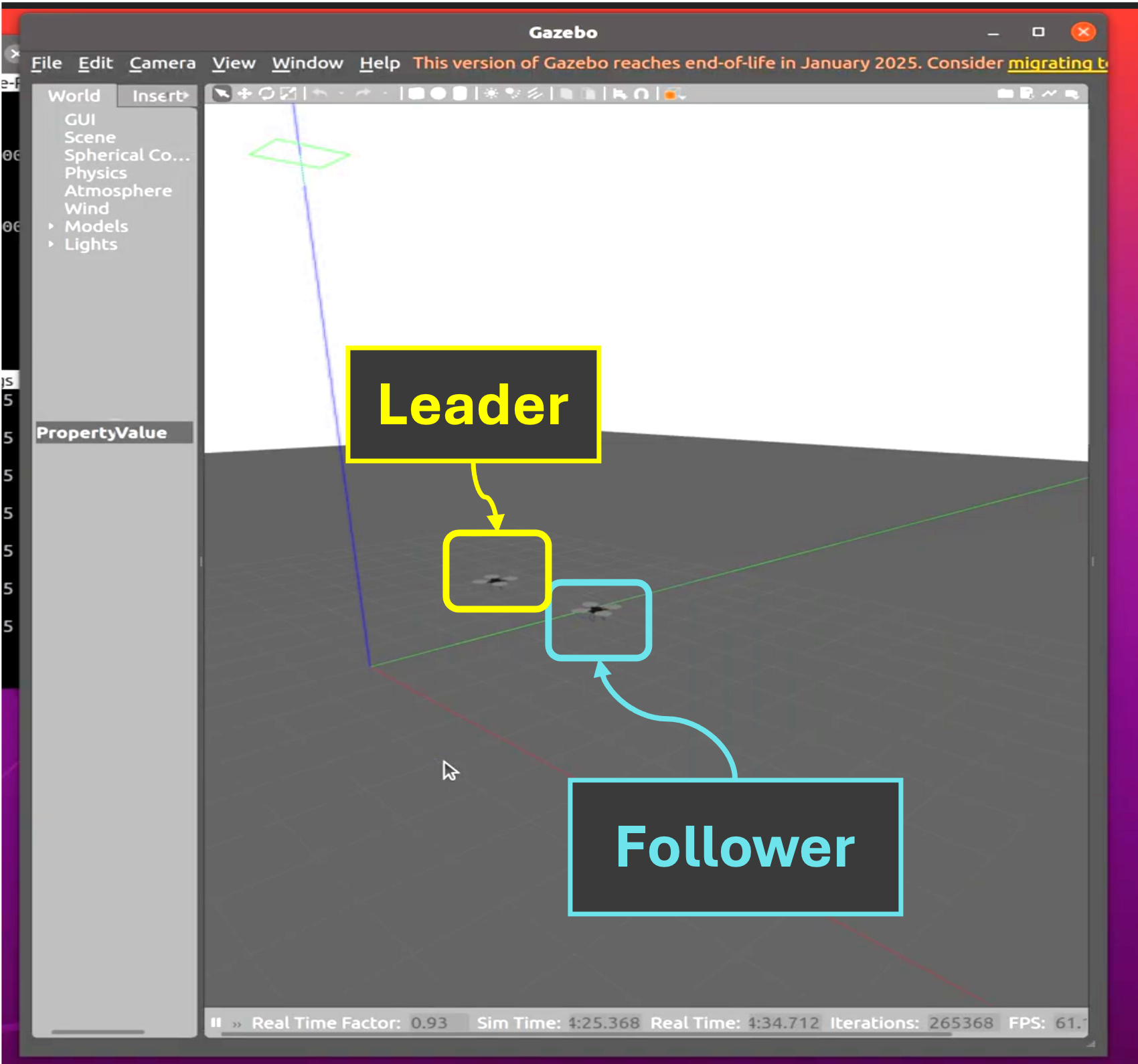}
    \caption{Robot-to-Robot Leader--Follower Coordination}
    \label{fig:robot-robot}
\end{subfigure}

\vspace{-1mm}
\caption{Representative 5G-NetBotSim: (a) human-to-robot teleoperation and (b) robot-to-robot leader-follower coordination.}
\label{fig:robotic_workload}
\vspace{-6mm}
\end{figure}

We evaluate 5G-NetBotSim using two representative connected-robot workloads: human-to-robot teleoperation and robot-to-robot leader--follower coordination, shown in Fig.~\ref{fig:robotic_workload}.

\subsubsection{Human-to-Robot Teleoperation}
For waypoint-based commands, Fig.~\ref{fig:latency_combined_top}, communication and application delays affect the task response time while the requested target remains unchanged. For velocity-based commands, newly received commands may interrupt the current trajectory and trigger replanning from the robot's current state. Thus, variations in command-delivery time can change the realized trajectory.

\subsubsection{Robot-to-Robot Leader--Follower Coordination}
\begin{figure*}[!t]
\centering

% ===================== Row 1: Low CQI =====================
\begin{subfigure}[t]{0.24\textwidth}
    \centering
    \includegraphics[width=0.8\columnwidth]{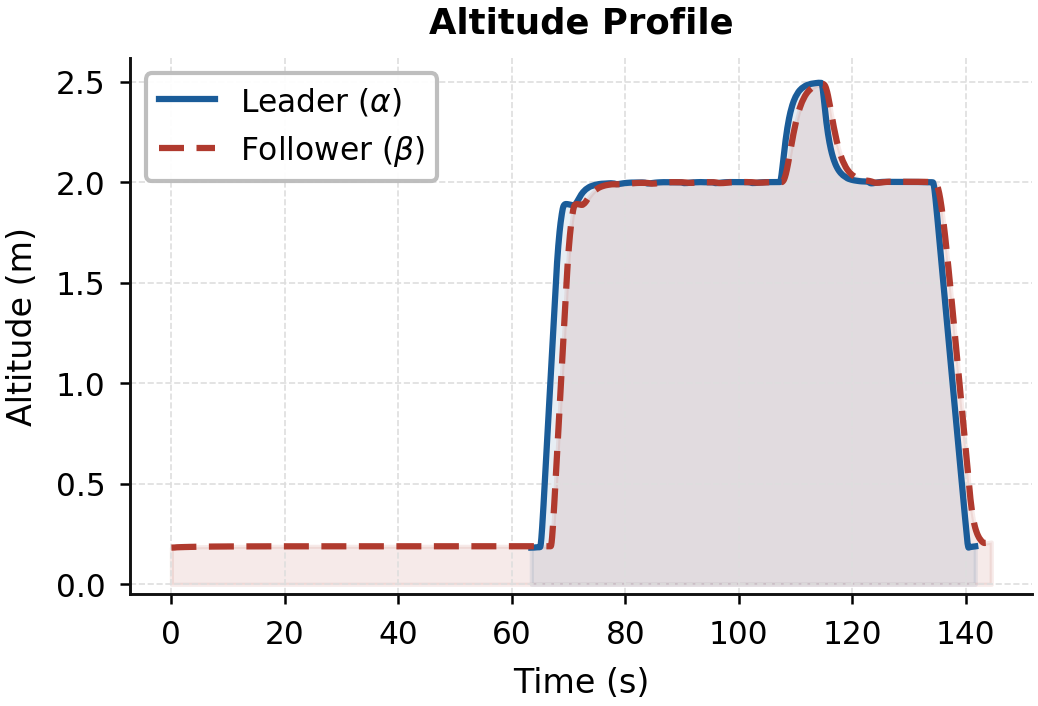}
    \caption{Low CQI: Altitude}
    \label{fig:altitude_low}
\end{subfigure}
\hfill
\begin{subfigure}[t]{0.24\textwidth}
    \centering
    \includegraphics[width=0.8\columnwidth]{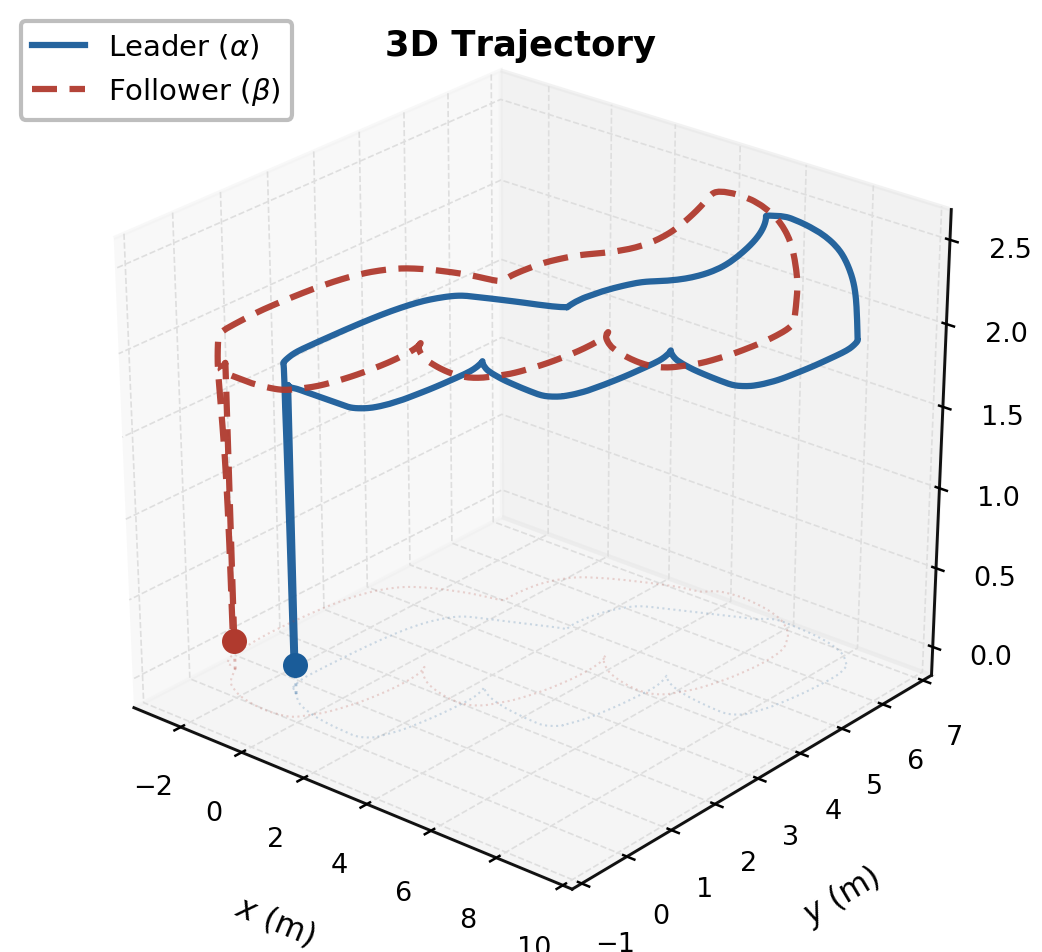}
    \caption{Low CQI: 3D trajectory}
    \label{fig:traj3d_low}
\end{subfigure}
\hfill
\begin{subfigure}[t]{0.24\textwidth}
    \centering
    \includegraphics[width=0.8\columnwidth]{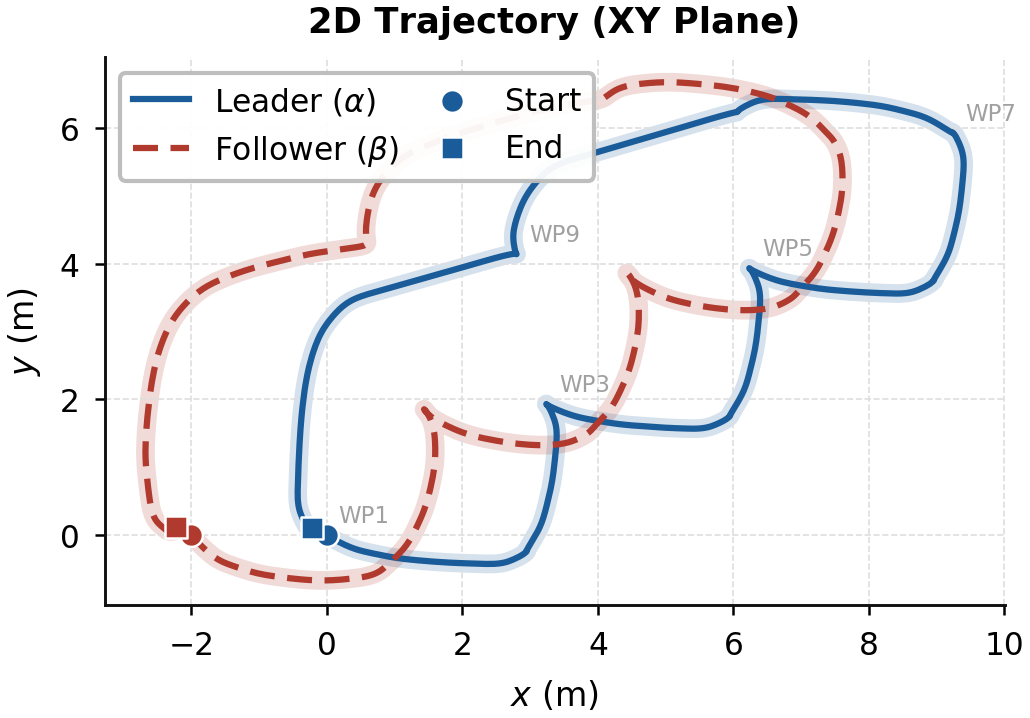}
    \caption{Low CQI: XY trajectory}
    \label{fig:traj2d_low}
\end{subfigure}
\hfill
\begin{subfigure}[t]{0.24\textwidth}
    \centering
    \includegraphics[width=0.8\columnwidth]{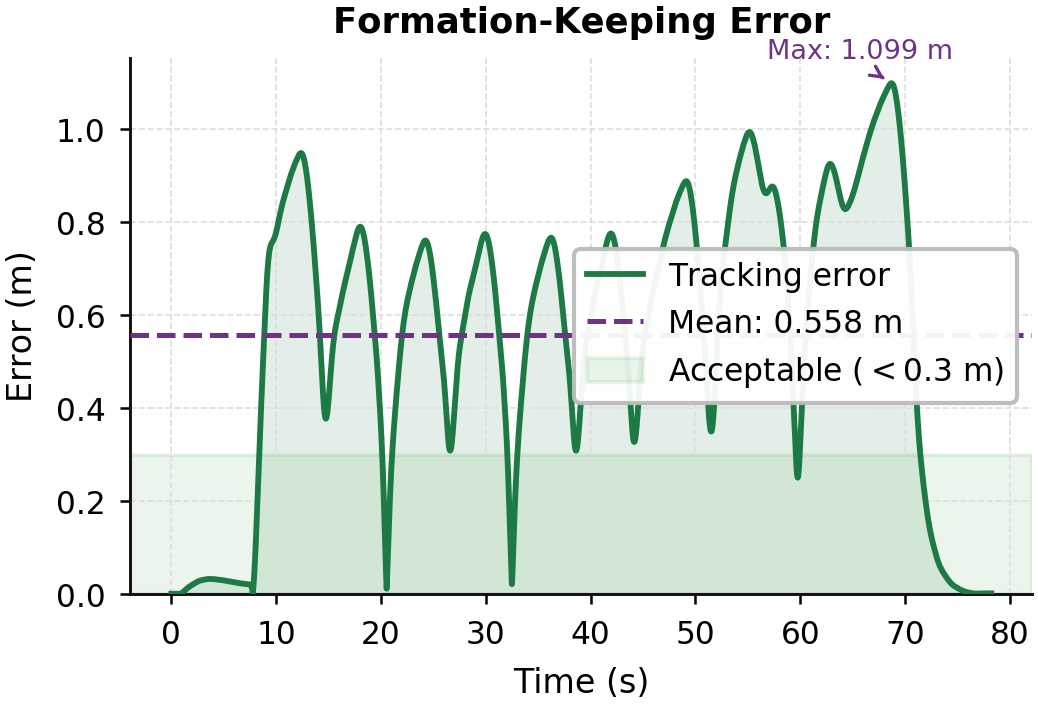}
    \caption{Low CQI: Tracking error}
    \label{fig:error_low}
\end{subfigure}

\vspace{1mm}

% ===================== Row 2: High CQI =====================
\begin{subfigure}[t]{0.24\textwidth}
    \centering
    \includegraphics[width=0.8\columnwidth]{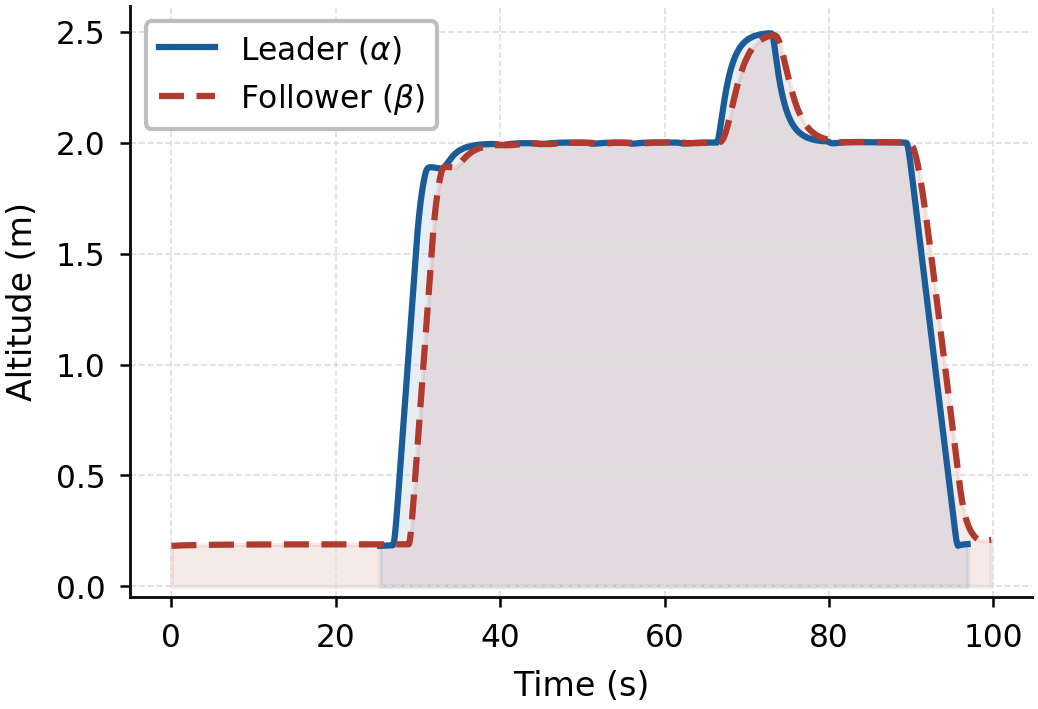}
    \caption{High CQI: Altitude}
    \label{fig:altitude_high}
\end{subfigure}
\hfill
\begin{subfigure}[t]{0.24\textwidth}
    \centering
    \includegraphics[width=0.8\columnwidth]{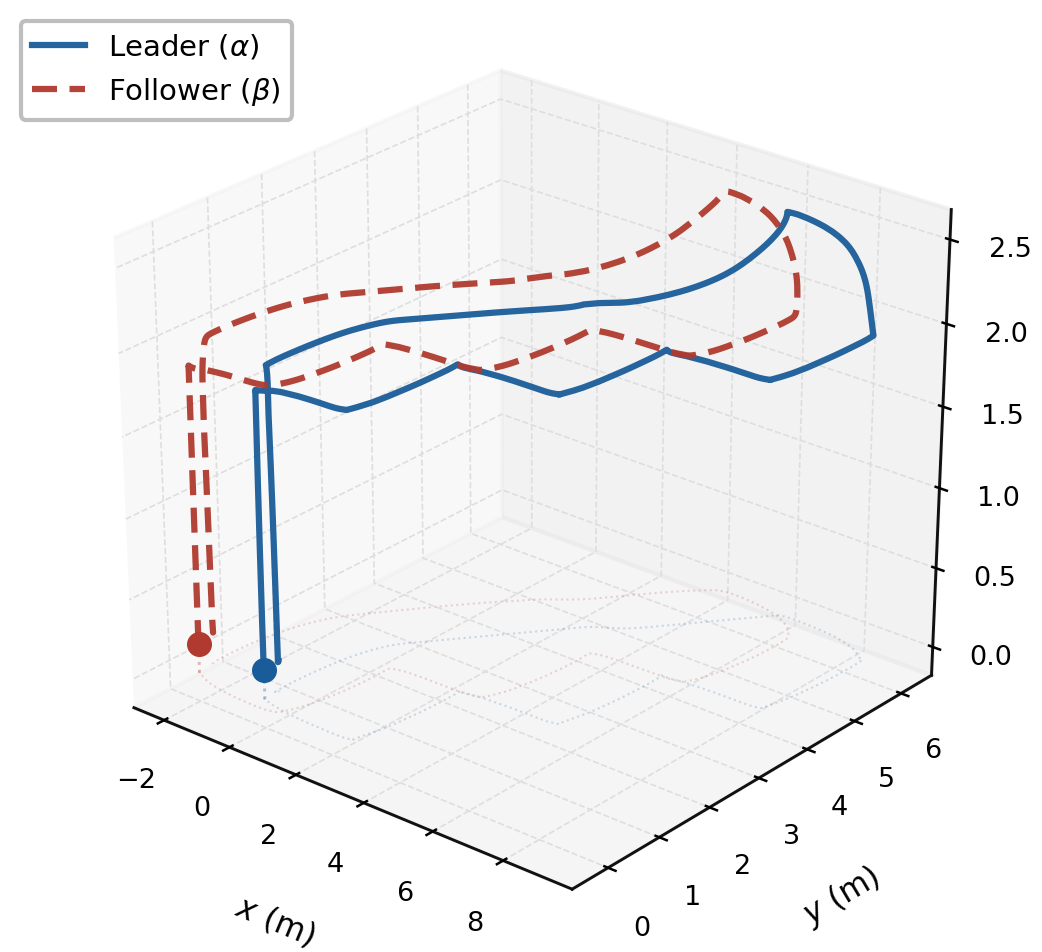}
    \caption{High CQI: 3D trajectory}
    \label{fig:traj3d_high}
\end{subfigure}
\hfill
\begin{subfigure}[t]{0.24\textwidth}
    \centering
    \includegraphics[width=\linewidth,height=2.4cm]{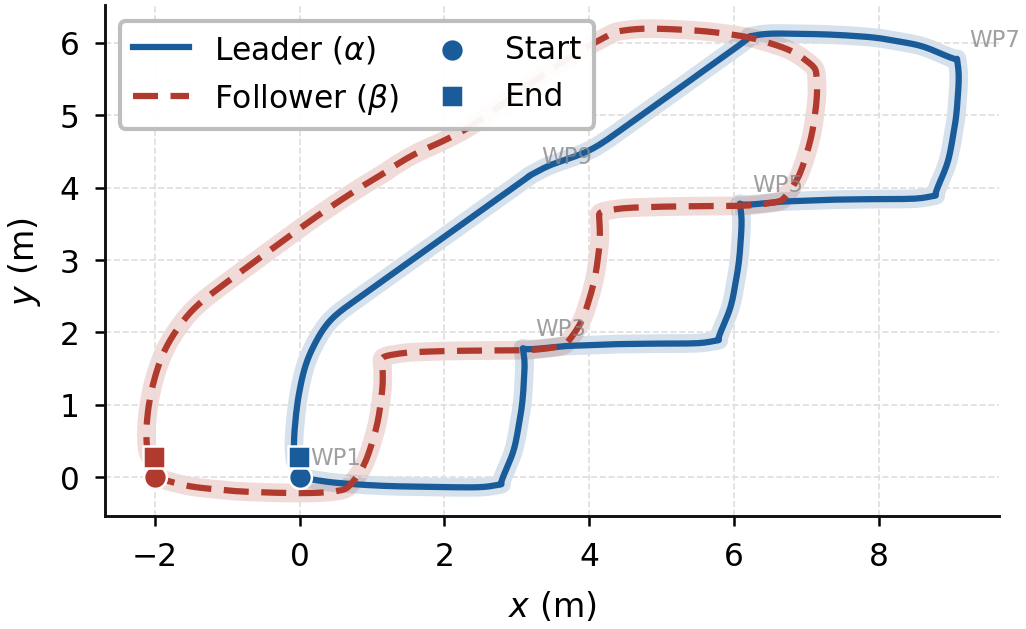}
    \caption{High CQI: XY trajectory}
    \label{fig:traj2d_high}
\end{subfigure}
\hfill
\begin{subfigure}[t]{0.24\textwidth}
    \centering
    \includegraphics[width=0.8\columnwidth]{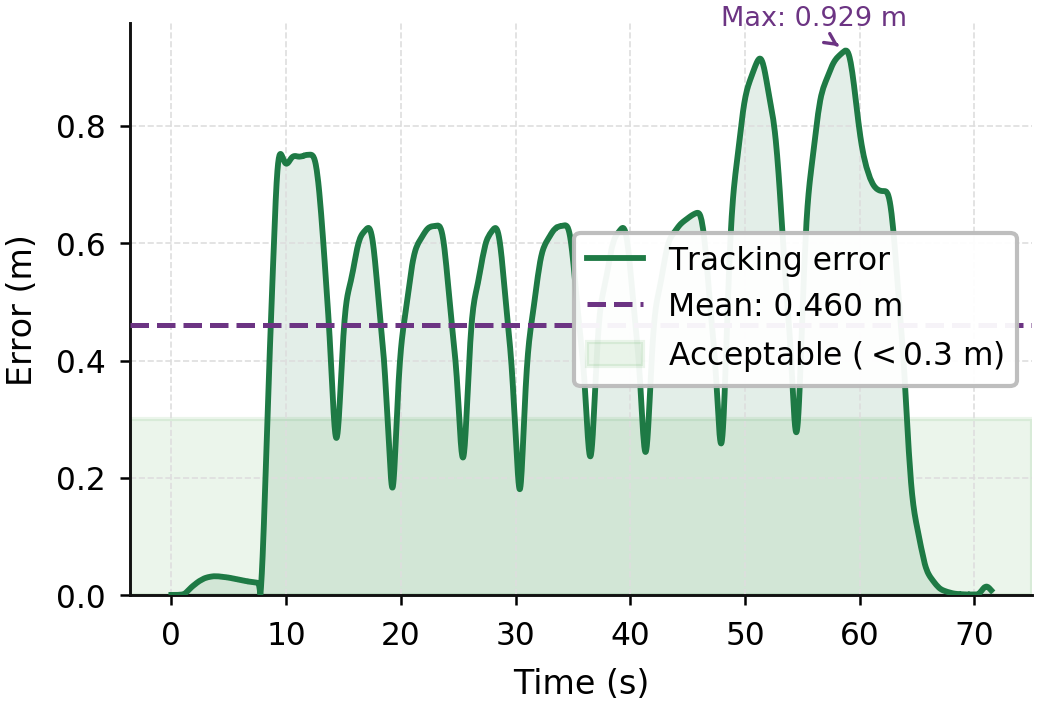}
    \caption{High CQI: Tracking error}
    \label{fig:error_high}
\end{subfigure}
\vspace{-1mm}
\caption{\small{Leader-follower trajectory and tracking performance under low- and high-CQI conditions. The first row shows the low-CQI case, and the second row shows the high-CQI case.}}
\label{fig:trajectory_comparison_cqi}
\vspace{-7mm}
\end{figure*}

The leader transmits its state to the follower at 10~Hz, and the follower uses the latest received state to maintain a desired 2~m offset. Under high CQI, (Fig.~\ref{fig:trajectory_comparison_cqi}), the mean tracking error decreases from 0.558~m to 0.460~m, corresponding to a 17.6\% reduction compared with low CQI. The improvement is most visible during turns and waypoint transitions, where delayed state updates increase the follower's spatial lag.

\section{Discussion}

The experiments demonstrate the value of protocol-stack-in-the-loop simulation when protocol behavior and its impact on robotic applications are of interest. Unlike abstract delay/loss models, an executable stack exposes effects such as CQI-dependent MCS selection, scheduling, per-UE behavior, and stage-level processing.
The current evaluation uses virtual-radio operation on a single host with two robot--UE pairs. Thus, the results characterize controlled protocol/application interactions rather than over-the-air fidelity or large-scale scalability. Physical-radio and larger multi-robot evaluations remain future work.

\section{Conclusion}

This paper presented 5G-NetBotSim, a protocol-stack-in-the-loop co-simulation framework integrating an executable srsRAN virtual-radio network with ROS/Gazebo robotic applications. Per-robot communication and ROS isolation together with stage-level timing allow network measurements to be related directly to task-facing robotic behavior.
Teleoperation and leader--follower experiments demonstrate how controlled channel-quality conditions propagate through MCS selection, throughput, communication latency, and robot execution. In the leader--follower experiment, improved channel quality reduces mean tracking error from 0.558~m to 0.460~m, corresponding to a 17.6\% improvement. Future work will investigate physical-radio validation, distributed execution, and larger multi-robot deployments.

% Reuse your existing bibliography file.
\bibliographystyle{IEEEtran}
\bibliography{ref}

\end{document}